\documentclass{aa}
\usepackage{graphicx, natbib,epsfig,amsmath,amssymb, mathrsfs, txfonts}
\DeclareGraphicsExtensions{.eps, .jpg, .ps}
\bibpunct{(}{)}{;}{a}{}{,}
\begin{document}

\title{Optimization of Apodized Pupil Lyot Coronagraph for ELTs}
\author{P. Martinez\inst{1}  \and A. Boccaletti\inst{1} \and M.
Kasper\inst{2} \and P. Baudoz\inst{1} \and C. Cavarroc\inst{1}}
\institute{LESIA, Observatoire de Paris Meudon, 5 pl. J. Janssen, 92195
Meudon, France \\
\email{patrice.martinez@obspm.fr}
\and European Southern Observatory, Karl-Schwarzschild-Strasse 2, D-85748, Garching, Germany}
\offprints{P. Martinez}

\abstract
{}
{We study the optimization of the Apodized Pupil Lyot Coronagraph (APLC) in the context of exoplanet imaging with ground-based telescopes. The APLC combines an apodization in the pupil plane with a small Lyot mask in the focal plane of the instrument. It has been intensively studied in the literature from a theoretical point of view, and prototypes are currently being manufactured for several projects. This analysis is focused on the case of Extremely Large Telescopes, but is also relevant for other telescope designs.} 
{We define a criterion to optimize the APLC with respect to telescope characteristics like central obscuration, pupil shape, low order segment aberrations and reflectivity as function of the APLC apodizer function and mask diameter. Specifically, the method was applied to two possible designs of the future European-Extremely Large Telescope (E-ELT).} 
{Optimum configurations of the APLC were derived for different telescope characteristics. We show that the optimum configuration is a stronger function of central obscuration size than of other telescope parameters. We also show that APLC performance is quite insensitive to the central obscuration ratio when the APLC is operated in its optimum configuration, and demonstrate that APLC optimization based on throughput alone is not appropriate.}
{}

\maketitle

\section{Introduction}
Over the past ten years many diffraction suppression systems were developed for direct detection of extrasolar planets. At the same time, promising ground-based projects were proposed and are currently under development like SPHERE at the VLT \citep{2006Msngr.125...29B} and GPI \citep{2006SPIE.6272E..18M}. Larger telescopes are desirable to improve performance of exoplanet searches towards lower masses and closer angular distances, ideally down to Earth-like planets. Several concepts of Extremely Large Telescopes (ELTs) are currently being studied all over the world (European-Extremely Large Telescope (E-ELT,  \citet{2004SPIE.5489..391D}), Thirty Meter Telescope (TMT, \citet{2006SPIE.6267E..71N}), Giant Magellan Telescope (GMT, \citet{2004SPIE.5489..441J})).

The characteristics of these telescope designs may have an impact on their high contrast imaging capabilities. Parameters like central obscuration, primary mirror segmentation, large spider arms, can impose strong limitations for many coronagraphs. It is therefore essential to indentify and evaluate a coronagraph concept which is well-suited to ELTs.

The Apodized Pupil Lyot Coronagraph (APLC) is one of the most promising concepts for ELTs. Its sensitivity to central obscuration is less critical than, e.g., for phase masks \citep{2000PASP..112.1479R, 2005ApJ...633.1191M} but the APLC still allows for a small inner working angle (IWA) and high throughput if properly optimized. Other amplitude concepts  (\citet[e.g.][]{2002ApJ...570..900K}) are also usable with centrally obscured aperture but suffer from low throughput especially if the IWA is small.
The potential of the APLC has already been demonstrated for arbitrary apertures \citep{2002A&A...389..334A, 2003A&A...397.1161S} and specific solutions for obscured apertures have been proposed \citep{2005ApJ...618L.161S}. 

In this paper, we analyze the optimization of the APLC and evaluate its sensitivity with respect to the main parameters mentioned above.

In section 2 we briefly revise the APLC formalism and we define a criterion for optimizing the coronagraph parameters. The impact of several telescope parameters on the optimal configuration is evaluated in section 3.
Then, section 4  shows an application of the APLC optimization to two potential ELTs designs. Finally, we derive conclusions.

\section{Apodization for centrally obscured pupil}
\subsection{Formalism}
In this section, we briefly revise the formalism of the APLC. The APLC is a combination of a classical Lyot coronagraph (hard-edged occulting focal plane mask, hereafter FPM) with an apodization in the entrance aperture. 
\begin{figure*}[!ht]
\includegraphics[width=9cm]{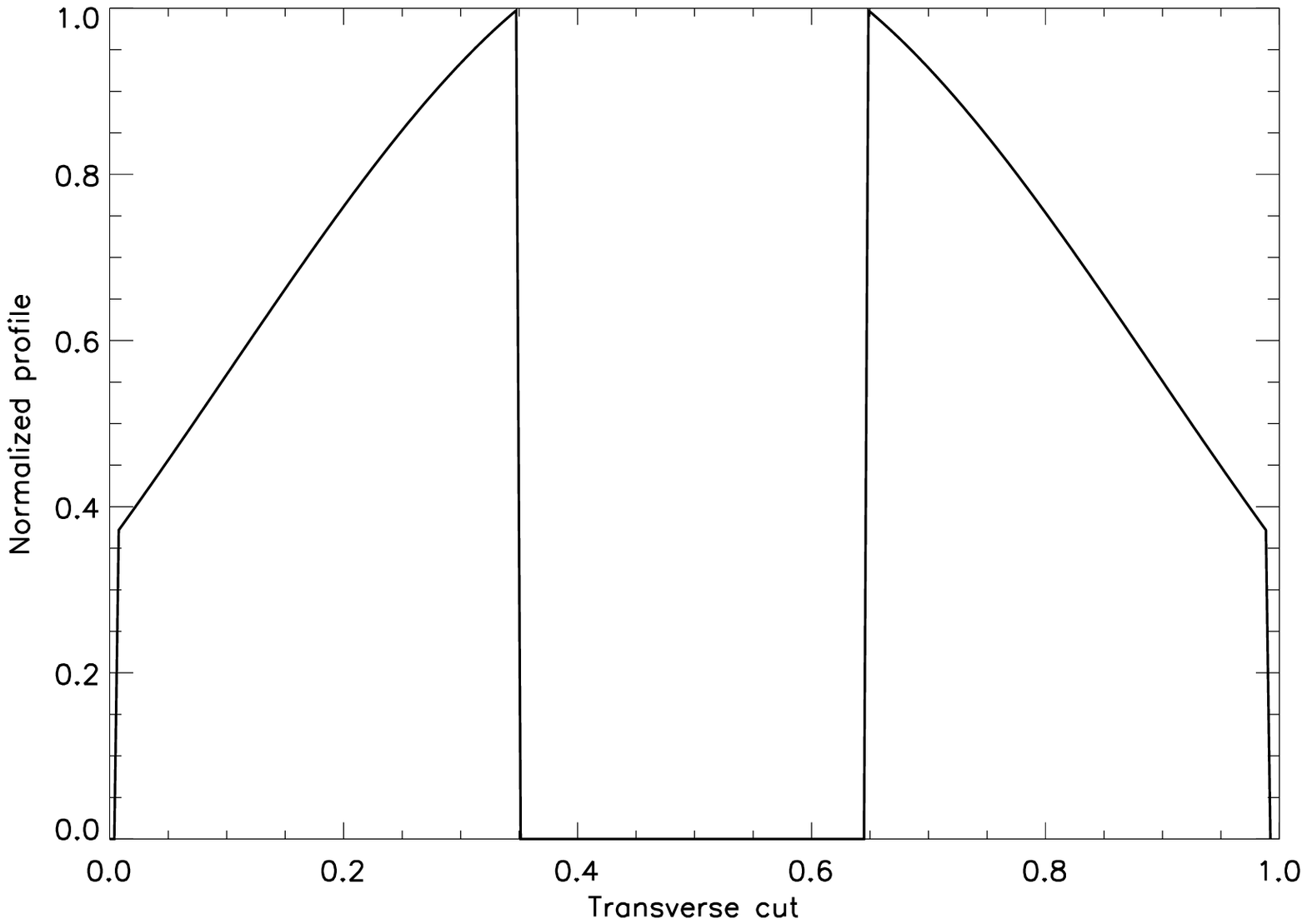}
\includegraphics[width=9cm]{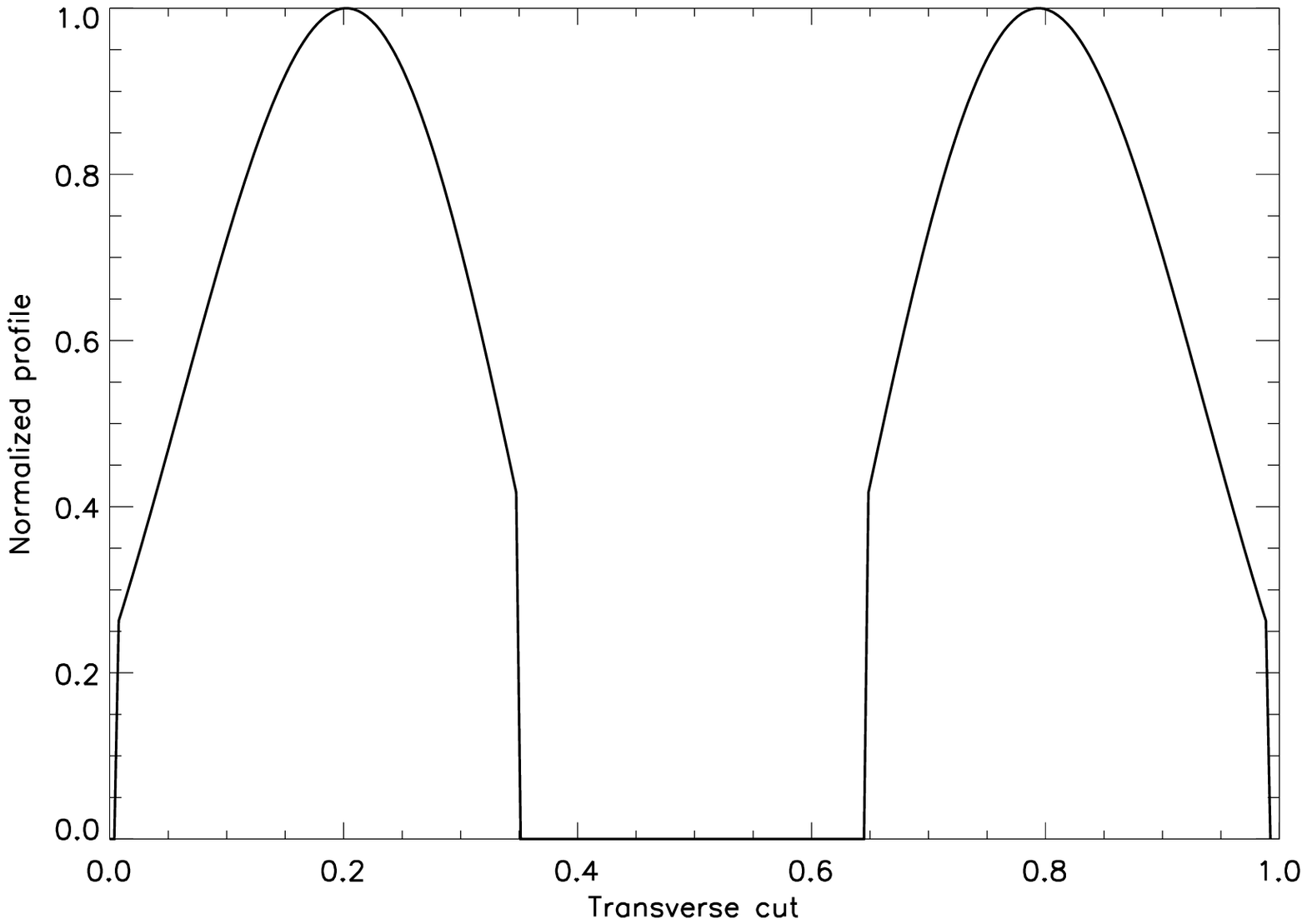}
\caption{Typical apodizer shape for the bell regime (left) and the bagel regime (right). Central obscuration is 30$\%$.}
\label{apodizershape1}
\end{figure*}

\begin{figure}[!ht]
\begin{center}
\includegraphics[width=9cm]{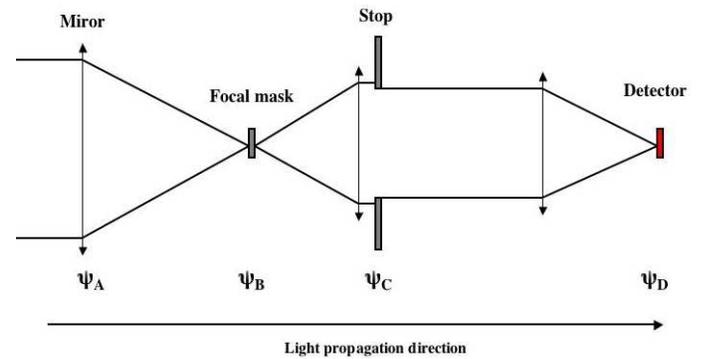}
\caption{Scheme of a coronagraph showing the pupil plane containing the apodizer ($\psi_A$), the focal plane with the FPM ($\psi_B$), the pupil image spatially filtered by the stop ($\psi_C$) and the detector plane ($\psi_D$).}
\label{coronoconcept}
\end{center}
\end{figure}

In the following, for sake of clarity, we omit the spatial coordinates $r$ and $\rho$ (respectively for pupil plane and focal plane). 
The function that describes the mask is noted $M$ (equal to 1 inside the coronagraphic mask and to 0 outside). With the mask absorption $\varepsilon$ ($\varepsilon = 1$ for an opaque mask),  the FPM is then equal to:
\begin{equation}
1 - \varepsilon M
\end{equation}
$P$ is the telescope aperture, and $\phi$ the profile of the apodizer. $\Pi$ describes the pupil stop function, which is considered -- in first approximation -- to be equal to the telescope aperture ($\Pi = P$). 
The coronagraphic process, corresponding to propagation from the telescope entrance aperture to the detector plane, is expressed in Eq. 2 to 6. Planes A, B, C and D correspond respectively to the telescope aperture, the coronagraphic focal plane, the pupil stop plane and the detector plane as defined in Fig. \ref{coronoconcept}. The Fourier transform of a function $f$ is noted $\hat{f}$. The symbol $\otimes$ denotes the convolution product. 
\noindent The entrance pupil is apodized in the pupil plane:
\begin{equation}
\psi_A =  P\phi
\end{equation}
 \noindent The complex amplitude of the star is spatially filtered (low frequencies) by the FPM:
\begin{equation}
\psi_B = \hat{\psi}_A \times [1 - \varepsilon M]
\end{equation}
 \noindent The exit pupil image is spatially filtered (high frequencies) by the stop:
\begin{equation}
\psi_C = \hat{\psi}_B \times \Pi
\end{equation}
\begin{equation}
\psi_C = [\psi_A - \varepsilon \psi_A \otimes \hat{M}] \times \Pi
\end{equation}
 \noindent The coronagraphic amplitude on the detector plane becomes:
\begin{equation}
\psi_D = \hat{\psi}_C = [\hat{\psi}_A  - \varepsilon \hat{\psi}_A M]  \otimes \hat{\Pi} 
\label{amplitudecorono}
\end{equation}

The coronagraphic process can be understood as a destructive interference between two waves (Eq. 5): the entrance pupil wave $P\phi$, noted $\psi_A$ and the diffracted wave by the mask (corresponding to $\varepsilon \psi_A \otimes \hat{M}$).
In the non-apodized case ($\phi=1$), the two wavefronts do not match each other, and the subtraction does not lead to an optimal starlight cancellation in the Lyot stop pupil plane.
A perfect solution is obtained if the two wavefronts are identical (i.e., the diffracted wave by the mask (\textit{M}) is equal to the pupil wave in amplitude). 
This latter case is obtained with the Apodized Pupil Phase Mask Coronagraph \citep{1997PASP..109..815R, 2002A&A...389..334A, 2003A&A...397.1161S}. For the APLC, the coronagraphic amplitude is minimized and proportional to the apodizer function. 

Considering a pupil geometry, the apodization function is related to the size of the FPM. More precisely, the shape of the apodizer depends on the ratio between the extent of $\hat{M}$ and the central obscuration size \citep{2005ApJ...618L.161S}. 
If the extent of $\hat{M}$ is bigger than the central obscuration, the apodizer takes a "bell" shape (typically it maximizes the transmission near the central obscuration of the pupil (see Fig.\ref{apodizershape1} (left) as illustration). On the contrary, if the extent of $\hat{M}$ is smaller than the central obscuration, the apodizer takes a "bagel" shape reducing transmission in the inner and outer part of the pupil (see Fig.\ref{apodizershape1} (right) as illustration). Thus, the apodizer shape depends on both,  the FPM  size and the central obscuration size.

Throughputs (apodizer transmission/pupil transmission) as a function of the FPM size is given in Fig. \ref{transmission} for different obscuration sizes (15 to 35 \%). These curves show a second maximum corresponding to the transition between the two apodizer regimes which depends on the central obscuration size. 

Since apodizer throughput does not evolve linearly with  FPM diameter, it is not trivial to determine the optimal  FPM/apodizer combination. Moreover, throughput might not be the only relevant parameter when optimizing a coronagraph.

A thorough signal to noise ratio analysis is definitely the right way to define the optimal  FPM/apodizer system, but this would be too instrument specific for the scope of this study.
Here, we investigate a general case for any telescope geometry and derive the corresponding optimal FPM size.

\subsection{APLC optimization criteria}
\begin{figure}[!ht]
\begin{center}
\includegraphics[width=9cm]{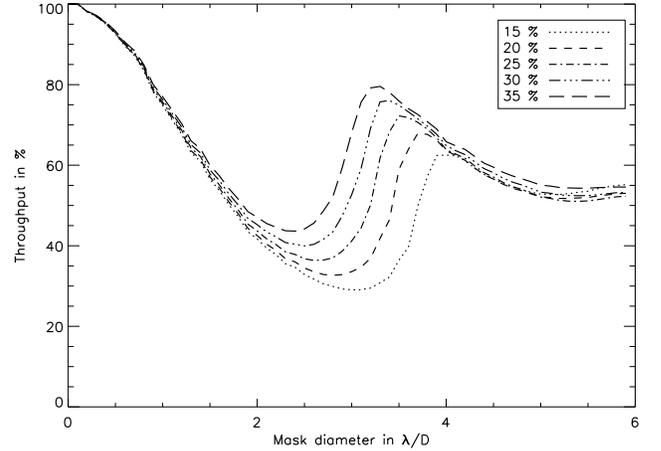}
\caption{Apodizer throughput (relative to full transmission of the telescope pupil) as a function of  FPM diameter for different obscuration sizes.}
\label{transmission}
\end{center}
\end{figure}

Usually, in Lyot coronagraphy, the larger the FPM diameter the larger the contrast. However, in the particular case of apodized Lyot coronagraph the transmission of an off-axis point-like object is not linear (Fig. \ref{transmission}) and then a trade-off has to be made between contrast and throughput.
This problem has been studied by \citet{2004EAS....12..165B} who evaluated optimal Lyot stops for any telescope pupil geometry and for any type of coronagraph. Based on this study, we propose a criterion adapted to the APLC to optimize the apodizer/ FPM combination. This criterion maximizes the coronagraphic performance while minimizing the loss of flux of the off-axis object. While not replacing a thorough signal-to-noise ratio evaluation, our criterion takes into account the modification of the off-axis PSF (in intensity and in shape) when changing the coronagraph parameters. 

Several metrics can be used to quantify the capability of a coronagraph (\citet[e.g.][]{2004EAS....12..165B}). Here, we use the contrast ($\mathscr{C}$) averaged over a range of angular radii : \\

\begin{equation}
\label{contrast}
\mathscr{C} = \frac{max\left(\mid \psi_D (\rho, \alpha)_{\varepsilon = 0} \mid^{2}\right)} {\left( \int^{2\pi}_{0} \int^{\rho_f}_{\rho_i} \mid \psi_D (\rho, \alpha)  \mid^{2}\rho\,  d\rho\, d\alpha \right) \slash \pi ({\rho_f}^{2} - {\rho_i}^{2})}  
\end{equation}

 \noindent where $\mathscr{C}$ is expressed in polar coordinates $\rho$ and $\alpha$. We denote by respectively $\rho_i$ and $\rho_f$ the short radii and the large radii, respectively, defining the area of calculation for $\mathscr{C}$. 
 
The attenuation of the off-axis object is given by the ratio of maximum image intensity with the apodizer only to the one without the coronagraph, i.e., without the apodizer and the FPM. This quantity differs from the throughput, since it also takes into account the modification of the PSF structure when changing the apodizer profile : 

\begin{equation}
max\left(\frac{\mid \psi_D (\rho, \alpha)_{\varepsilon = 0} \mid^{2} }{\mid \hat{P}(\rho, \alpha) \mid^{2}}\right)
\label{psfattenuation}
\end{equation}

 \noindent Now, let us define the criterion $C_{\mathscr{C}}$ as the product of $\mathscr{C}$ and Eq. \ref{psfattenuation}. \\
  \begin{equation}
 C_{\mathscr{C}} = \mathscr{C} \times max\left(\frac{\mid \psi_D (\rho, \alpha)_{\varepsilon = 0} \mid^{2} }{\mid \hat{P}(\rho, \alpha) \mid^{2}}\right)
 \label{CC}
\end{equation}

 \noindent The first term of $C_{\mathscr{C}}$ (Eq.  \ref{contrast}, which characterizes the performances of the coronagraphic system) is then adapted to the region of interest in the coronagraphic image and can be well matched to the instrument parameters. 
 
 \noindent The second term (Eq. \ref{psfattenuation}) takes into account the modification of the PSF structure when changing the apodizer profile and guarantees a reasonably moderate attenuation of the off-axis PSF maximum intensity (i.e, guarantees that when the coronagraph rejects the star it does not reject the planet as well). 

Although our criterion cannot replace a thorough signal-to-noise ratio analysis (no instrumental model, no noise terms), it presents a reasonable approach by assuming the residual light leaking through the coronagraph as noise. Our criterion allows us to investigate the trade-off between performance and throughput while keeping the study general and independent of a specific instrument setup.

Moreover, the validity of this criterion is supported by the pupil stop optimization study of \citet{2004EAS....12..165B} who was facing a problem similar to ours, and also by the results presented and discussed afterwards in this paper.

\section{Sensitivity analysis}
\subsection{Assumptions}
Based on the previously defined criterion, we now analyze the behavior of several telescope parameters as a function of the size of the FPM (and hence APLC characteristics) with the main objective to explore possibilities how to optimize the APLC configuration for a given ELT design. 
One advantage of $C_{\mathscr{C}}$ is that the area of optimization in the focal plane can be well matched to the instrumental parameters. For that reason, we have limited the search area and investigated $C_{\mathscr{C}}$ only between $\rho_i = 3 \lambda/D$ at small radii and $\rho_f =100 \lambda/D$ at large radii.
These limits correspond to the Inner Working Angle (distance at which an off-axis object reaches a significant transmission) and to the high-order Adaptive Optics (AO) cut-off frequency, respectively. At radii larger than the AO cut-off frequency, the coronagraph will only have a minor effect since atmospheric turbulence is not corrected and atmospheric speckles dominate.

For the simulations presented in the next sections, we assume a circular pupil with 30\% central obscuration. The central obscuration ratio is left as a free parameter only in section \ref{sec:obscuration} where we evaluate its impact.
The pupil stop is assumed identical to the entrance pupil including spider arms \citep{2005ApJ...633..528S}. 
Section \ref{sec:spider}, where the impact of the spider arms' size is analyzed, assumes 42-m telescope. Elsewhere, simulation results do not depend on the telescope diameter.
Apodizer profiles were calculated numerically with a Gerchberg-Saxton iterative algorithm \citep{GSalgo}. The pixel sampling in the focal plane is 0.1 $\lambda/D$, and the pupil is sampled with 410 pixels in diameter. When phase aberrations are considered we are adopting a wavelength of 1.6$\mu$m corresponding to the H-band in the near infrared.
\begin{center}
\begin{table*}
\centering
\caption{Optimum  FPM diameter (and hence APLC characteristics) for several obscuration sizes and criteria.}
\label{obscuration}
\begin{tabular}{ccc|cc}
\hline \hline
&\multicolumn{2}{c|}{$C_{\mathscr{C}}$ }  & \multicolumn{2}{c}{Max. throughput} \\
\cline{2-5} \\
Obstruction size (\%) &  FPM ($\lambda/D$) & Throughput ($\%$) &  FPM ($\lambda/D$) & Throughput ($\%$) \\
\hline
10         &  4.3  &  59.4&    4.1  & 62.2 \\
15 	&  4.3  & 58.3 &    4.0   &62.4  \\ 
20 	&  4.4  & 55.8&   3.8   & 65.5 \\
25 	&  4.6  & 52.7&    3.6   & 67.9\\
30 	&  4.7 & 51.2&    3.5   & 68.7\\
35 	&  4.9  & 49.4&    3.3   & 70.4 \\
\hline
\end{tabular}
\end{table*}
\end{center}
\begin{figure}[!ht]
\begin{center}
\includegraphics[width=9cm]{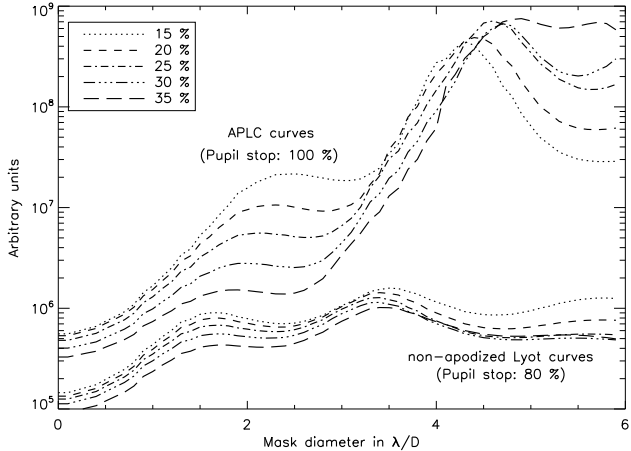}
\caption{$C_{\mathscr{C}}$ average between 3 and 100 $\lambda/D$ as a function of the FPM diameter and obscuration sizes, in the case of the APLC and classical Lyot coronagraph.}
\label{centralobscuration1}
\end{center}
\end{figure}

\subsection{Critical parameter impacts}
In the following sub-sections, we are studying the impact of 2 major categories of diffraction effects. The first category deals with amplitude variations: central obscuration, spider arms, primary mirror segmentation, segment-to-segment reflectivity variation, and pupil shear (misalignment of the coronagraph stop with respect to the instrument pupil). Inter-segment gaps and other mechanical secondary supports are not considered, since they would require finer pixel sampling in the pupil image, resulting in  prohibitively large computation times with a non-parallel computer.
In addition, some mechanical secondary supports can be much smaller than the main spider arms. To first approximation, their effects can be considered to be similar to the ones produced by spider arms.

The second category is related to phase aberrations, that we assumed to be located in the pupil plane (no instrumental scintillation). We only modeled low-order segment aberrations (piston, tip-tilt, defocus, astigmatism). Higher orders are less relevant for the optimization of the FPM size, but can have a significant impact on the coronagraphic performance. 

The amplitude diffraction effect of gaps is partially accounted for (at least for infinitely small gaps) by the phase transition we are generating between primary mirror segments.
\begin{figure}[!ht]
\begin{center}
\includegraphics[width=9cm]{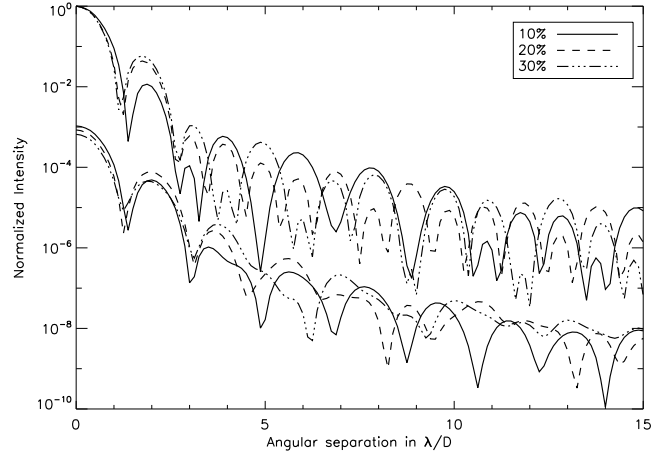}
\caption{Radial profiles of PSFs and coronagraphic images obtained with optimal APLC (using $C_{\mathscr{C}}$) for several obscuration sizes.}
\label{centralobscuration2}
\end{center}
\end{figure}
\subsubsection{Central obscuration}
\label{sec:obscuration}

\noindent The first parameter we are evaluating is the central obscuration. High contrast instruments have to deal with central obscuration ratios which typically range from 10\% to 35\% (CFHT: 35\%, HST: 33\%, VLT: 14 \%). ELTs will likely have larger obscurations than current 8-m class telescopes to preserve a reasonable size of the telescope structure.
In Fig. \ref{centralobscuration1}, the criterion $C_{\mathscr{C}}$ is shown for different obscuration sizes ranging from 10 to 35 \%. The curves show two maxima. The first one is located near 2 $\lambda$/D and experiences a large contrast variation while the second one (near 4$\lambda$/D) shows a smaller dispersion.
\begin{figure}[!ht]
\begin{center}
\includegraphics[width=8cm]{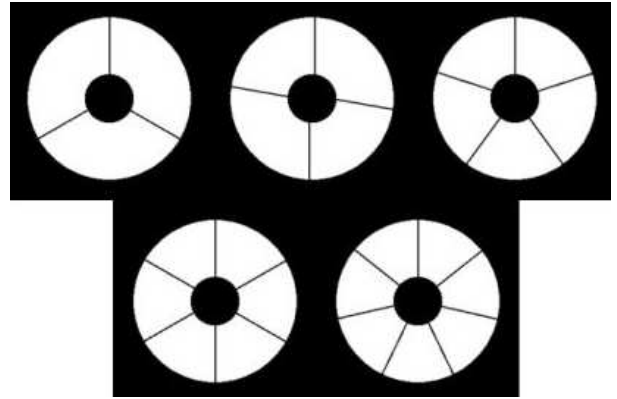}
\caption{Pupil configurations considered here.}
\label{spider1}
\end{center}
\end{figure}
Table 1 summarizes these results and gives the position of the second maximum versus the obscuration size for the criterion previously mentioned and for a criterion based solely on the maximum throughput (like in Fig. \ref{transmission}).

\begin{figure}[!ht]
\begin{center}
\includegraphics[width=9cm]{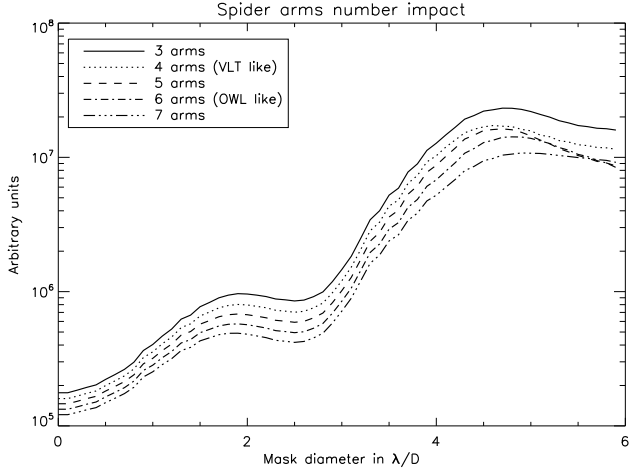}
\caption{$C_{\mathscr{C}}$ average between 3 and 100 $\lambda/D$ as a function of the FPM diameter and number of spider arms. Spider thickness is set to 62 cm.}
\label{spider2}
\end{center}
\end{figure}
\begin{figure}[!ht]
\begin{center}
\includegraphics[width=9cm]{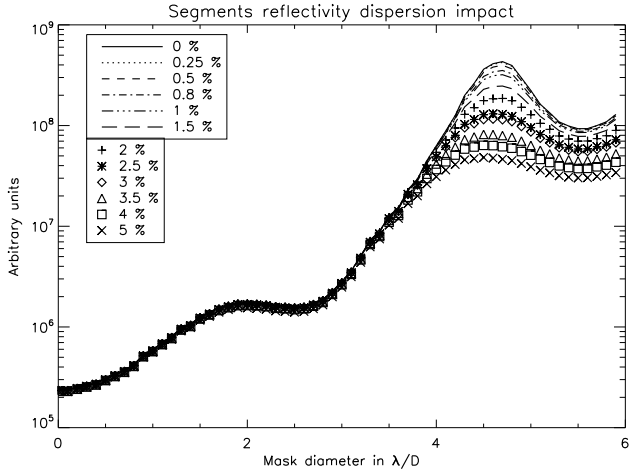}
\caption{$C_{\mathscr{C}}$ average between 3 and 100 $\lambda/D$ as a function of the FPM diameter and reflectivity variations.}
\label{reflectivity}
\end{center}
\end{figure}

If we only consider the second maximum, which is more promising in terms of contrast and appears less sensitive, the optimal FPM diameter ranges from 4.3 to 4.9 $\lambda/D$ for obscuration ratios between 10 to 35\%. 
Here, our criterion $C_{\mathscr{C}}$ is more relevant than just throughput, since it is better adapted to the region of interest in the coronagraphic image and to the modification of the PSF structure. 
We see a non-linear increase of optimum FPM size with the obscuration ratio because more starlight is redistributed in the Airy rings of the PSF. A solely throughput-based consideration shows the opposite behavior with a larger dispersion of the FPM size, which is not consistent with the effect on the PSF structure. However, at small obscuration sizes (10\%-15\%), maximum throughput yields a similar optimal FPM diameter as $C_{\mathscr{C}}$.

\begin{figure}[!ht]
\begin{center}
\includegraphics[width=9cm]{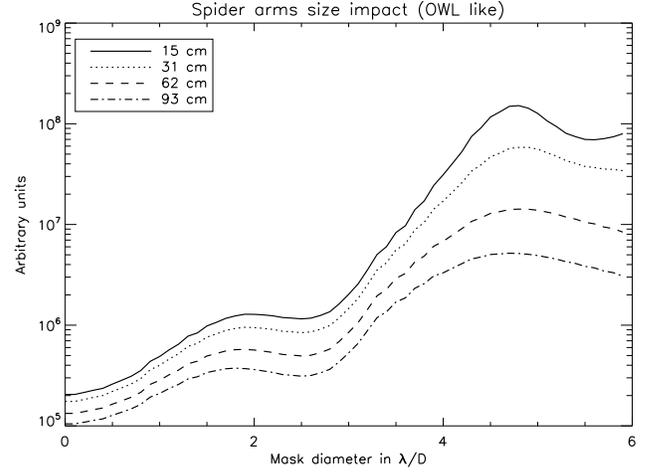}
\caption{$C_{\mathscr{C}}$ average between 3 and 100 $\lambda/D$ as a function of the FPM diameter and spider arms thickness. Number of spider arms is set to 6.}
\label{spider3}
\end{center}
\end{figure}
\begin{figure}[!ht]
\begin{center}
\includegraphics[width=9cm]{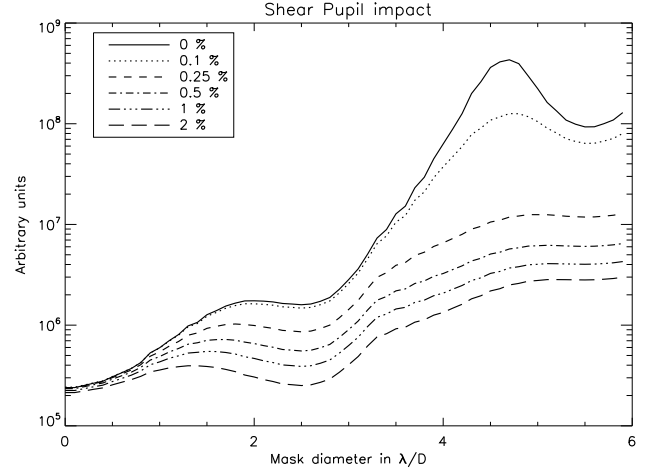}
\caption{$C_{\mathscr{C}}$ average between 3 and 100 $\lambda/D$ as a function of the FPM diameter and pupil shear.}
\label{shearpupil}
\end{center}
\end{figure}
We consider this result as evidence for the relevance of our criterion $C_{\mathscr{C}}$ to optimize the FPM size (and hence the APLC characteristics) with respect to the size of the central obscuration. Moreover, the validity of our criterion is also supported by the comparison of coronagraphic PSFs using an optimized APLC in Fig. \ref{centralobscuration2}. The optimized APLC allows for a contrast performance which is rather insensitive to the central obscuration size.

\subsubsection{Spider arms}
\label{sec:spider}
On an ELT, the secondary mirror has to be supported by a complex system of spider arms ($\sim$ 50 cm) and cables ($\sim$ 30-60 mm) to improve stiffness. 
Evaluating the influence of these supports is important in the context of coronagraphy. 

The pixel sampling of our simulations limited by available computer power does not allow us to model the thinnest mechanical supports. However, the impact of these supports on the PSF structure will be similar to the one of spider arms but at a reduced intensity level. 
Several configurations were considered as shown in Fig.\ref{spider1}. 
As the number of spider arms increases from 3 to 7, the contrast gets worse (but no more than a factor of 2). The curves in Fig.\ref{spider2} are almost parallel, indicating that the number of spider arms has no significant influence on the optimal FPM size. The second maximum of $C_{\mathscr{C}}$ peaks at 4.7 $\lambda/D$ with a small dispersion of 0.2 $\lambda/D$. 
\begin{table*}
\caption{APLC optimization for an obscuration of 30\% }
\label{}
\centering
\begin{tabular}{lcc}
\hline \hline
Parameters & Value range  & Optimal APLC configuration ( FPM range in $\lambda/D$) \\
\hline
Obscuration & 30\% & 4.7  \\
Spider (arm) & 3 - 7 & 4.6 - 4.8  \\
Spider (size) & 15 - 90 cm & 4.6 - 4.8  \\
Shear pupil &  0.5 - 2 \% & 4.7 - 4.9 \\
Segment reflectivity  & 0.25 - 5 \% & 4.5 - 4.7 \\
Low-order aberrations & 1 - 100 nm rms & 3.5 - 6.0 \\
Chromatism ($\Delta \lambda \slash \lambda$) & 1.4 - 5 \%  &  4.7 - 4.8 \\
Chromatism ($\Delta \lambda  \slash \lambda $)  & 5 - 20 \%  &  4.8 - 5.3 \\
\hline
\end{tabular}
\end{table*}
\begin{figure}[!ht]
\begin{center}
\includegraphics[width=9cm]{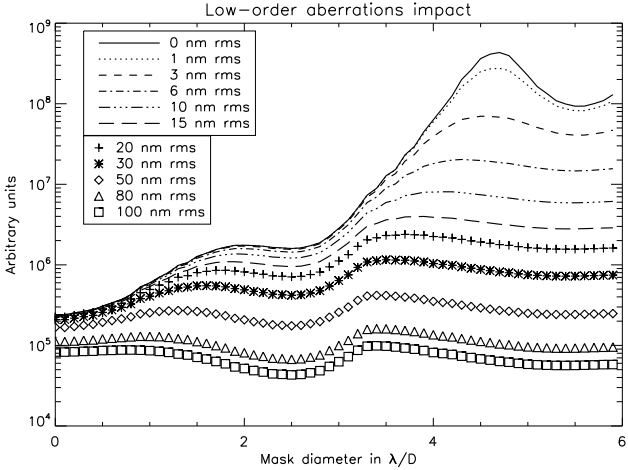}
\caption{$C_{\mathscr{C}}$ average between 3 and 100 $\lambda/D$ as a function of the FPM diameter and low-order aberrations.}
\label{aberrations}
\end{center}
\end{figure}
Assuming a 6-spider arms configuration (OWL-like), we also analyzed the sensitivity to spider arms thickness from 15 cm to 93 cm (Fig.\ref{spider3}). 
The increasing width of the spider arms tends to flatten the profile of $C_{\mathscr{C}}$, making the selection of an optimal FPM more difficult (or less relevant) for very large spider arms. 
However, for the actual size of spider arms likely being of the order of 50 cm, the optimal size of the FPM (and hence APLC) is still 4.7 $\lambda/D$.
\subsubsection{Segments reflectivity variation}

\noindent The primary mirror of an ELT will be segmented because of its size, and a potential resulting amplitude effect is segment-to-segment reflectivity variation. 
We show the APLC optimization sensitivity for segment reflectivity variation from 0 to 5 \% peak-to-valley in Fig.\ref{reflectivity}. For this simulation, the primary mirror was assumed to consists of $\sim$750 hexagonal segments.

The criterion $C_{\mathscr{C}}$ is robust for FPMs smaller than 4 $\lambda/D$. A loss of performance with reflectivity variation is observed for larger FPM. However, the optimal FPM size remains located at 4.7 $\lambda/D$ with a small dispersion of 0.2 $\lambda/D$.

\subsubsection{Pupil shear}
As already mentioned above an APLC includes several optical components : apodizer, FPM and pupil stop.
The performance of the APLC also depends on the alignment of these components. In particular, the pupil stop has to accurately match the telescope pupil image. This condition is not always satisfied, and the telescope pupil may undergo significant mismatch which could amount to more than 1\% of its diameter. 
The pupil shear is the mis-alignment of the pupil stop with respect to the telescope pupil image. It is an issue especially for ELTs for which mechanical constraints are important for the design. For example, the James Webb Space Telescope is expected to deliver a pupil image for which the position is known at about 3-4\%. Therefore, the performance of the mid-IR coronagraph \citep{2004EAS....12..195B} will be strongly affected. 
\begin{figure}[!ht]
\begin{center}
\includegraphics[width=9cm]{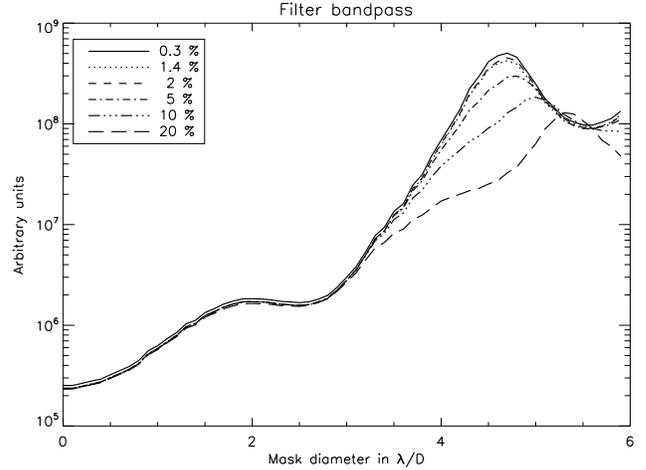}
\caption{$C_{\mathscr{C}}$ average between 3 and 100 $\lambda/D$ as a function of FPM diameter and the filter bandpass.}
\label{chromatism}
\end{center}
\end{figure}
\begin{table}[!ht]
\begin{center}
\caption{Chromatism effects synthesis}
\label{spectralR}
\begin{tabular}{ccccc}
\hline
\hline
$\Delta \lambda \slash \lambda$ (\%) &  $FPM (\lambda/D)$ & $FPM_{\lambda_{max}}$ ($\lambda/D$) & $F_1$ & $F_2$ \\
\hline
0.3 & 4.70 & 4.70 &  1.0 & 1.0 \\
1.4 & 4.70  & 4.73 & 1.1 & 1.1 \\
2  & 4.70 & 4.75 & 1.1 &  1.1 \\
5  & 4.80  & 4.82 & 1.6 & 1.6 \\
10 & 5.00 & 4.94 & 2.6 & 3.7 \\
20 & 5.30 & 5.20 & 3.7 & 14.6 \\
50  & 5.90  & 5.87 & 26.3 & 180.9 \\
\hline
\end{tabular}
\end{center}
\end{table} 
On SPHERE, the planet-finder instrument for the VLT (2010), the pupil shear was identified as a major issue and a dedicated Tip-Tilt mirror was included in the design to preserve the alignment at a level of 0.2\% \citep{2006tafp.conf..353B}.

The behavior of $C_{\mathscr{C}}$ in Fig. \ref{shearpupil} is somewhat different from the behavior of the previous parameters. The loss of performance is significant even for small FPM. However, the criterion is still peaking at 4.7 $\lambda/D$ with a variation of about 0.2 $\lambda/D$ although above 4.5 $\lambda/D$ the curves are rather flat indicating that a larger FPM would not improve performance.

\subsubsection{Static aberrations}
\begin{figure*}[!ht]
\begin{center}
\includegraphics[width=5cm]{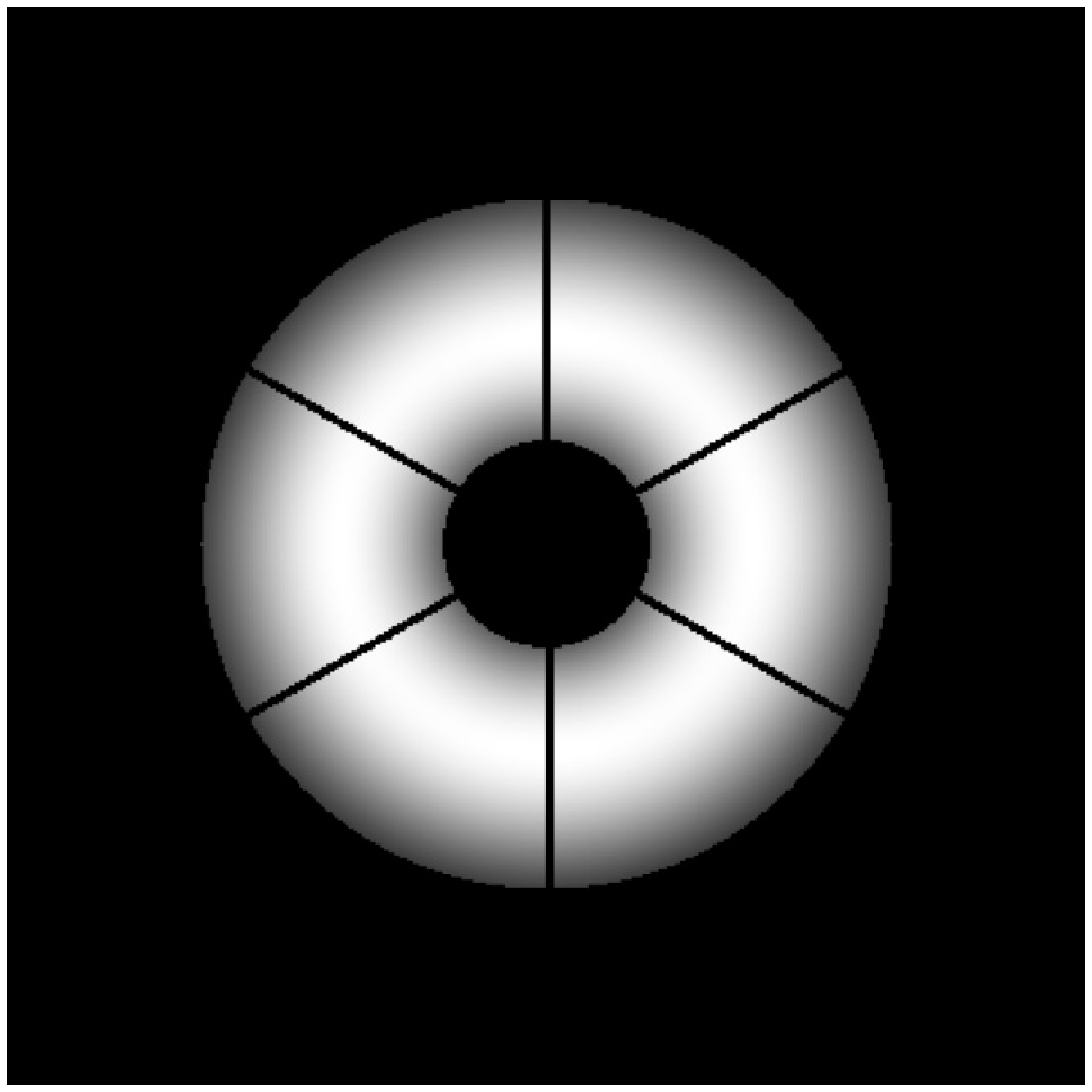}
\includegraphics[width=5cm]{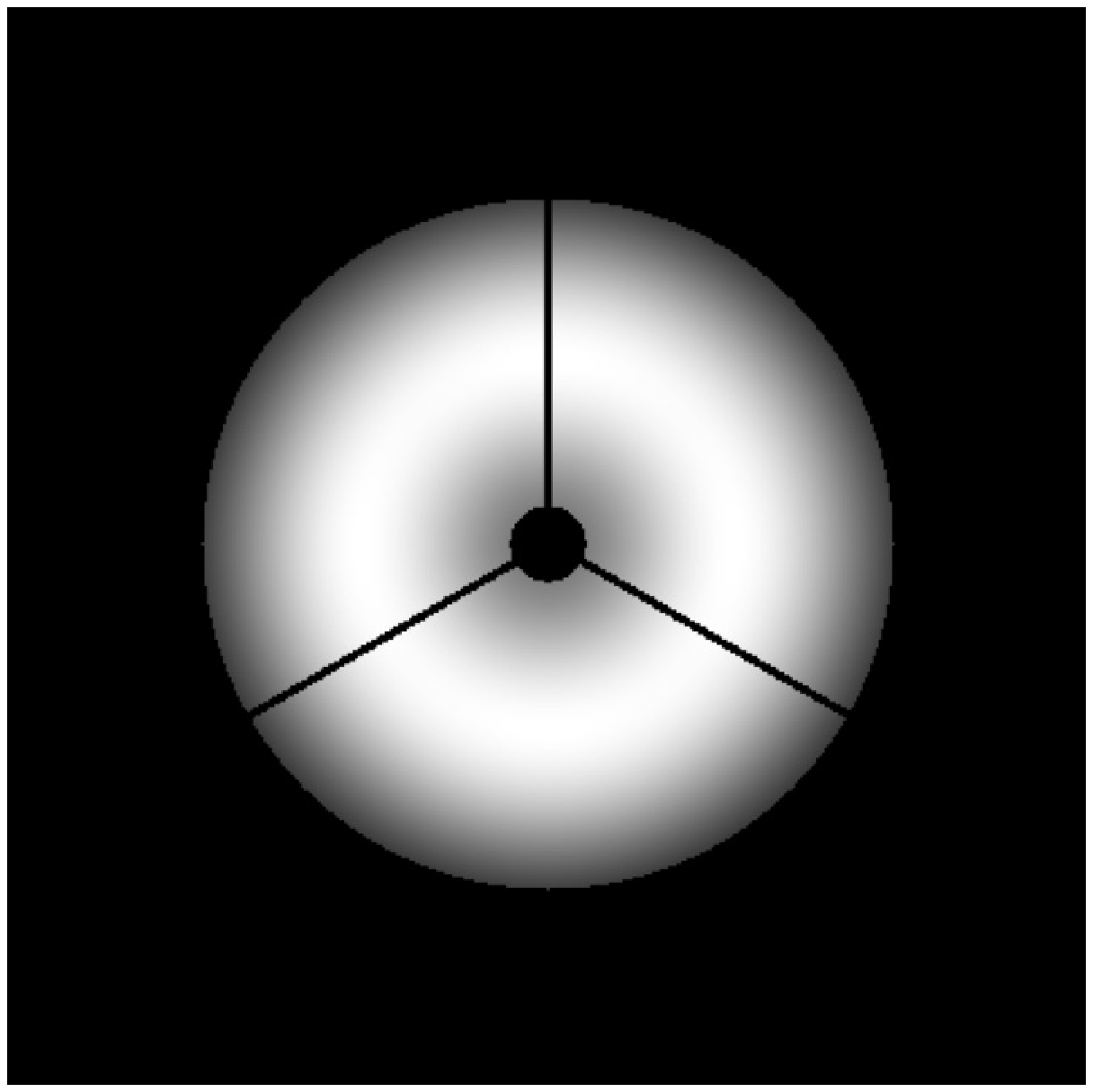}
\caption{Optimized apodized E-ELT apertures: telescope design 1 (left), telescope design 2 (right).}
\label{design}
\end{center}
\end{figure*}

\noindent Here, static aberrations refer to low-order aberrations on the segments of the large primary mirror. We separately investigated the effect of piston, tip-tilt, defocus and astigmatism,  and found the behavior to be similar for all these aberrations.
In contrast to the other defects, both, the performance and the optimal FPM diameter (optimal APLC) are very sensitive to low-order aberrations.

As the amplitude of aberrations increases, the dependency of $C_\mathscr{C}$ on FPM diameter becomes flatter and the optimal FPM size is getting smaller (Fig. \ref{aberrations}). A larger FPM would not decrease performance enormously. For values larger than 15nm, there is no longer clear evidence of an optimal size beyond $\sim 3.5 \lambda/D$. The performance is rather insensitive to the actual FPM size.

Even though low-order aberrations strongly affect APLC performance, their presence has virtually no impact on the optimized configuration. The fairly constant performance in the presence of larger low-order aberrations indicates that low-order aberrations are not a relevant parameter for the optimization of the APLC.

\subsubsection{Chromatism}

\noindent All previous analysis was performed for monochromatic light of the wavelength $\lambda_0$. However, as with the classical Lyot coronagraph, the APLC performance should depend on the ratio between FPM size and PSF size and therefore on wavelength. Hence, the impact of chromatism on the APLC optimization must be evaluated. We note that the chromatism of the APLC can also be mitigated by a slight modification of the standard design  \citep{2005PASP..117.1012A}.

Figure \ref{chromatism} and Tab. \ref{spectralR} present the results of the simulations for several filter bandpass widths ($\Delta \lambda \slash \lambda$) when using the standard monochromatic APLC. As long as the filter bandpass is smaller than 5 \%, the optimal FPM size and performance are nearly the same as in the monochromatic case.

The values displayed in columns 4 and 5 of Tab. \ref{spectralR} quantify the loss of contrast due to chromaticity with respect to the monochromatic case for the APLC being optimized to the filter bandpass ($F_{1}$) and to the central wavelength of the band ($F_{2}$). These two factors begin to differ significantly from each other at a filter bandpass larger than 5 \%. Hence, optimization of the APLC for chromatism is needed for a filter bandpass exceeding this value. 

An efficient way of optimizing an APLC for broad band application is to optimize it for the longest wavelength of the band, which leads to results that are within 0.1$\lambda/D$ of the true optimal FPM size. This behavior can be explained by the non-symmetrical evolution of the residual energy in the coronagraphic image around the optimal FPM size at $\lambda_0$ \citep{2003A&A...397.1161S}. Another way to minimize chromaticity would be to calculate the apodizer profile for the central wavelength and only optimize the FPM diameter considering the whole bandpass. We compared the behavior of both methods for $\Delta \lambda \slash \lambda = 20\%$: they are actually very comparable in terms of performance. 

\section{Application to the E-ELT}
\label{sec:Application}

In this section, we apply the tools and results from the APLC optimization study discussed in the previous section to the two telescope designs proposed for the European-ELT. The objective is to confirm our optimization method and to produce contrast idealized profiles which admittedly must not be confused with the final achievable contrast in the presence of a realistic set or instrumental error terms.

\subsection{Starting with telescope designs} 
We assume a circular monolithic primary mirror of 42 meters diameter. Segmentation errors are not taken into account, although we note that the E-ELT primary mirror consists of hexagonal segments with diameters ranging from 1.2 to 1.6 meters in its current design.

There are two competing telescope designs considered: a 5 mirror arrangement (design 1) and a 2 mirror Gregorian (design 2). For our purpose, the two designs differ by their central obscuration ratios and the number of spider arms. Design 1 (Fig.\ref{design} left) is a 30\% obscured aperture with 6 spider arms of 50 cm 
and design 2 (Fig.\ref{design} right) is a 11\% obscured aperture with 3 spider arms of 50 cm 
These numbers are likely to be subject to change as the telescope design study is progressing. Mechanical supports (non-radial cables of the secondary mirror support) and intersegment gaps are not considered for the reasons mentioned in section discussing spider arms. 

In such conditions and taking into account the previous sensitivity analysis on central obscuration, spider arms, and chromatism ($\Delta \lambda \slash \lambda = 20 \%$) we found an optimal APLC configurations with the apodizer designed for 4.8 and 4.3 $\lambda/D$ and with a FPM size of 5 and 4.3 $\lambda/D$ for design 1 and 2, respectively. 
\begin{figure*}[!ht]
\begin{center}
\includegraphics[width=9cm]{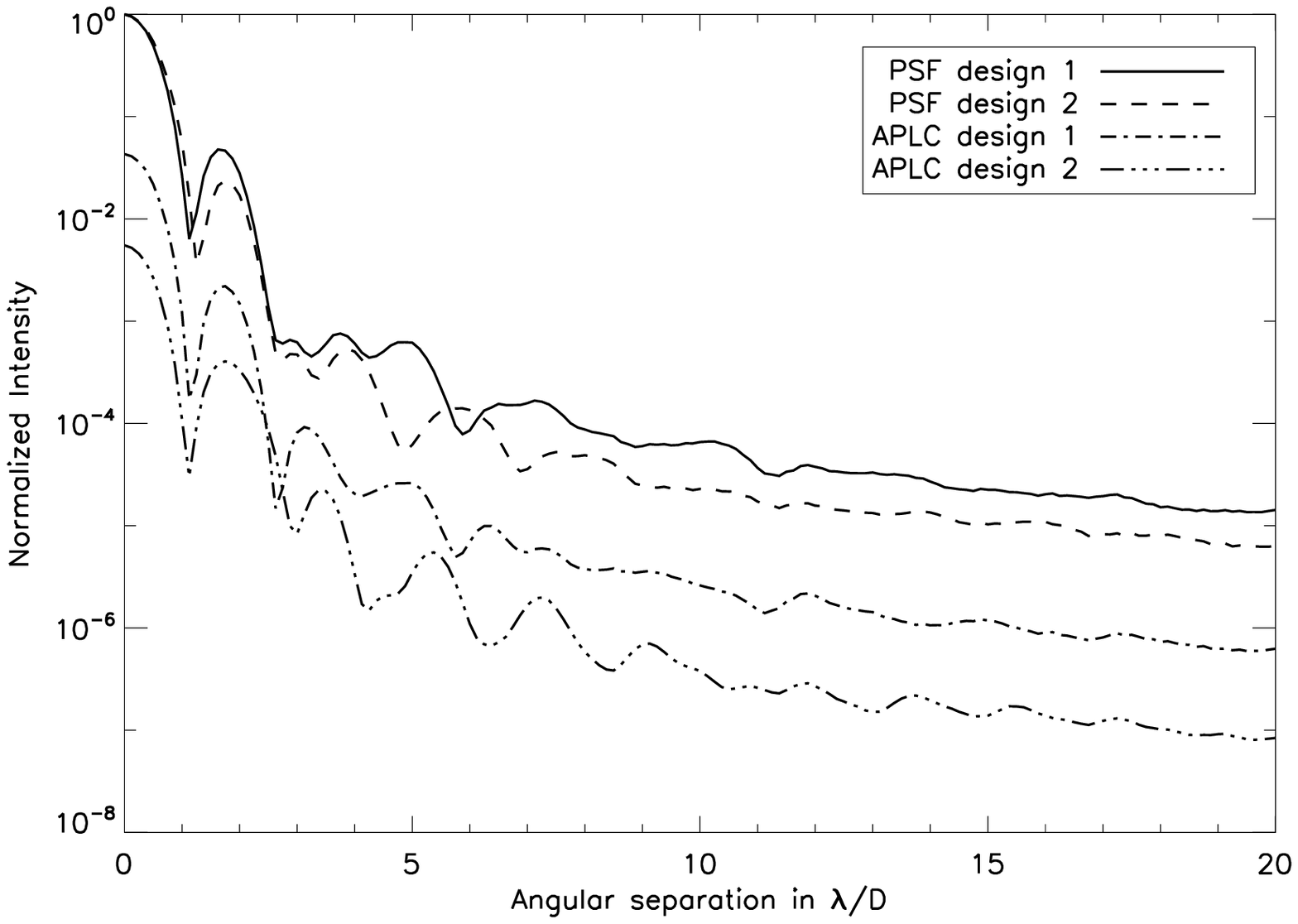}
\includegraphics[width=9cm]{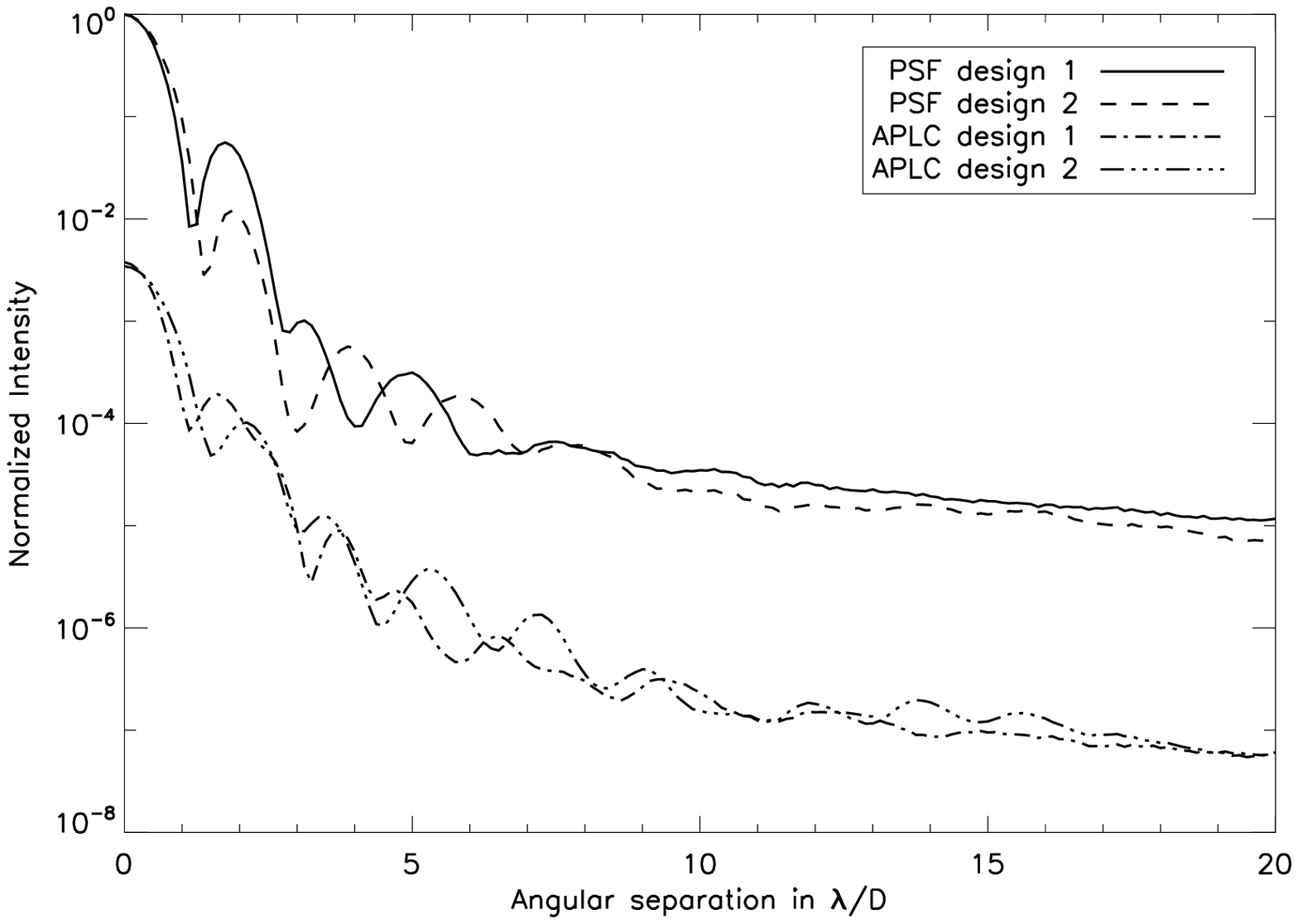} 
\caption{Radial profiles of PSFs and coronagraphic images ($\lambda/\Delta \lambda = 5$) for the 2 designs considering throughput optimization (up) or $C_\mathscr{C}$ optimization (bottom).}
\label{PSFcomp}
\end{center}
\end{figure*}

\citet{2005ApJ...633..528S} has demonstrated that optimization or under-sizing of the pupil stop is not necessary with the APLC. We independently verified and confirm this result using our criterion applied on the stop rather than on the mask.

\subsection{Radial contrast}
\label{sec:perfELT}
As already shown in section \ref{sec:obscuration}, the optimal APLC configuration with our criterion is different to the optimal configuration considering throughput as a metric. We can now demonstrate this difference using contrast profiles. 
Figure \ref{PSFcomp} compares the coronagraphic profiles based on throughput optimization (apodizer and FPM size are 3.5 and 4.1 $\lambda/D$ for design 1 and 2, see Figure \ref{transmission}) with the one obtained from optimization with our criterion. 

For design 2, the optimization with both methods leads to similar APLC configurations (4.3 and 4.1 $\lambda/D$). Hence, the contrast performance between them differs by only a factor of 3. For design 1, instead, the gain by using our criterion for the optimization is a staggering factor 10 in contrast. In addition, the plot shows that APLC contrast performance only weakly depends on the telescope geometry with this optimization method. This is an important result, which means that the APLC can efficiently cope with a large variety of telescope designs.


\section{Conclusion}
The Apodized Pupil Lyot Coronagraph is believed to be a well suited coronagraph for ELTs and to the search of extrasolar planets with direct imaging. The high angular resolution of such large telescopes relaxes the constraints on the Inner Working Angle (IWA) of a coronagraph which is an important issue for high contrast imaging instruments on 8-m class telescopes. Hence, coronagraphs with a relatively large IWA like the APLC present an interesting alternative to the small IWA coronagraphs such as the phase mask coronagraphs. 

The objective of this paper was to analyze the optimization of APLC in the context of ELTs. We defined a criterion ($C_\mathscr{C}$) similar to the one use by \citet{2004EAS....12..165B} for the general problem of Lyot stop optimization in coronagraphy. We then analyzed the behavior of this criterion as a function of the FPM diameter in the presence of different telescope parameters. The optimal FPM is determined by the maximum value of the criterion. A sensitivity analysis was carried out for the several telescope parameters like central obscuration, spiders, segment reflectivity, pupil shear, low-order static aberrations and chromatism. 
Some of these parameters are not relevant for APLC optimization such as low-order aberrations which provide a pretty flat response of the criterion to FPM diameter when applied at reasonably large amplitudes. However, ELTs are not yet defined well enough to predict the level of static aberrations coronagraphs will have to deal with.

The parameter which had the largest impact on the optimum FPM diameter is the central obscuration. An obscuration ratio of 30\% leads to and optimal APLC of 4.7 $\lambda/D$. In most cases, the optimal sizes derived for other telescope parameters are quite consistent with the one imposed by the central obscuration. The dispersion of the FPM size is no larger than 0.2$\lambda$/D given the range of parameters we have considered. We also demonstrated that APLC optimization based on throughput alone is not appropriate and leads to optimal FPM sizes that are decreasing with increasing obscuration ratios. This behavior is opposite to the one derived using our criterion. The superior quality of our criterion is supported by the comparison of contrast profiles obtained with both optimization methods in sections \ref{sec:perfELT} and \ref{sec:obscuration}.

Although the idealized simulations presented in this paper do not consider atmospheric turbulence and instrumental defects, they allow us to find the optimal APLC configuration and PSF contrast for each case. \citet{2006A&A...447..397C} show that the ultimate contrast achievable by differential imaging (speckle noise suppression system to enhance the contrast, \citet{1999PASP..111..587R, 2000PASP..112...91M, 2003PASP..115.1363B, 2004ApJ...615..562G}) with a perfect coronagraph is not sensitive to atmospheric seeing but critically depends on static phase and amplitude aberrations. Our results therefore present the possibility to extend this study to the more realistic case of a real coronagraph taking into account relevant effects releated to telescope properties.

In addition, we have also started a development of APLC prototypes whose characteristics were defined with the present numerical analysis. Experiments with these prototypes will be carried out during the next year in the near IR on the High Order Test-bench \citep{2006SPIE.6272E..81V} developed at the European Southern Observatory. The practical study of the APLC will also benefit from prototyping activities led by the department of Astrophysics at the University of Nice (LUAN) and carried out for development of SPHERE for the VLT. 

\begin{acknowledgements}
P. M would thanks Pierre Riaud for helpful discussions. This activity is supported by the European Community under its Framework Programme 6, ELT Design Study, Contract No. 011863.
\end{acknowledgements}

\nocite{*}
\bibliography{MyBiblio}

\begin{thebibliography}{25}
\expandafter\ifx\csname natexlab\endcsname\relax\def\natexlab#1{#1}\fi

\bibitem[{{Aime}(2005)}]{2005PASP..117.1012A}
{Aime}, C. 2005, \pasp, 117, 1012

\bibitem[{{Aime} {et~al.}(2002){Aime}, {Soummer}, \&
  {Ferrari}}]{2002A&A...389..334A}
{Aime}, C., {Soummer}, R., \& {Ferrari}, A. 2002, \aap, 389, 334

\bibitem[{{Andersen} {et~al.}(2003){Andersen}, {Ardeberg}, {Beckers},
  {Goncharov}, {Owner-Petersen}, {Riewaldt}, {Snel}, \&
  {Walker}}]{2003SPIE.4840..214A}
{Andersen}, T., {Ardeberg}, A.~L., {Beckers}, J., {et~al.} 2003, in Presented
  at the Society of Photo-Optical Instrumentation Engineers (SPIE) Conference,
  Vol. 4840, Future Giant Telescopes. Edited by Angel, J. Roger P.; Gilmozzi,
  Roberto. Proceedings of the SPIE, Volume 4840, pp. 214-225 (2003)., ed.
  J.~R.~P. {Angel} \& R.~{Gilmozzi}, 214--225

\bibitem[{{Baba} \& {Murakami}(2003)}]{2003PASP..115.1363B}
{Baba}, N. \& {Murakami}, N. 2003, \pasp, 115, 1363

\bibitem[{{Beuzit} {et~al.}(2006{\natexlab{a}}){Beuzit}, {Feldt}, {Dohlen},
  {Mouillet}, {Puget}, {Antici}, {Baruffolo}, {Baudoz}, {Berton}, {Boccaletti},
  {Carbillet}, {Charton}, {Claudi}, {Downing}, {Feautrier}, {Fedrigo}, {Fusco},
  {Gratton}, {Hubin}, {Kasper}, {Langlois}, {Moutou}, {Mugnier}, {Pragt},
  {Rabou}, {Saisse}, {Schmid}, {Stadler}, {Turrato}, {Udry}, {Waters}, \&
  {Wildi}}]{2006Msngr.125...29B}
{Beuzit}, J.-L., {Feldt}, M., {Dohlen}, K., {et~al.} 2006{\natexlab{a}}, The
  Messenger, 125, 29

\bibitem[{{Beuzit} {et~al.}(2006{\natexlab{b}}){Beuzit}, {Mouillet}, {Moutou},
  {Dohlen}, {Fusco}, {Puget}, {Udry}, {Gratton}, {Schmid}, {Feldt}, {Kasper},
  \& {The Vlt-Pf Consortium}}]{2006tafp.conf..353B}
{Beuzit}, J.~L., {Mouillet}, D., {Moutou}, C., {et~al.} 2006{\natexlab{b}}, in
  Tenth Anniversary of 51 Peg-b: Status of and prospects for hot Jupiter
  studies, ed. L.~{Arnold}, F.~{Bouchy}, \& C.~{Moutou}, 353--355

\bibitem[{{Boccaletti}(2004)}]{2004EAS....12..165B}
{Boccaletti}, A. 2004, in EAS Publications Series, ed. C.~{Aime} \&
  R.~{Soummer}, 165--176

\bibitem[{{Boccaletti} {et~al.}(2004){Boccaletti}, {Riaud}, {Baudoz},
  {Baudrand}, {Reess}, \& {Rouan}}]{2004EAS....12..195B}
{Boccaletti}, A., {Riaud}, P., {Baudoz}, P., {et~al.} 2004, in EAS Publications
  Series, ed. C.~{Aime} \& R.~{Soummer}, 195--204

\bibitem[{{Cavarroc} {et~al.}(2006){Cavarroc}, {Boccaletti}, {Baudoz}, {Fusco},
  \& {Rouan}}]{2006A&A...447..397C}
{Cavarroc}, C., {Boccaletti}, A., {Baudoz}, P., {Fusco}, T., \& {Rouan}, D.
  2006, \aap, 447, 397

\bibitem[{{Dierickx} {et~al.}(2004){Dierickx}, {Brunetto}, {Comeron},
  {Gilmozzi}, {Gont{\'e}}, {Koch}, {le Louarn}, {Monnet}, {Spyromilio},
  {Surdej}, {Verinaud}, \& {Yaitskova}}]{2004SPIE.5489..391D}
{Dierickx}, P., {Brunetto}, E.~T., {Comeron}, F., {et~al.} 2004, in Presented
  at the Society of Photo-Optical Instrumentation Engineers (SPIE) Conference,
  Vol. 5489, Ground-based Telescopes. Edited by Oschmann, Jacobus M., Jr.
  Proceedings of the SPIE, Volume 5489, pp. 391-406 (2004)., ed. J.~M.
  {Oschmann}, Jr., 391--406

\bibitem[{{Gerchberg} \& {Saxton}(1972)}]{GSalgo}
{Gerchberg}, R.~W. \& {Saxton}, W.~O. 1972, in Optik 35, 237-246

\bibitem[{{Guyon}(2004)}]{2004ApJ...615..562G}
{Guyon}, O. 2004, \apj, 615, 562

\bibitem[{{Johns} {et~al.}(2004){Johns}, {Angel}, {Shectman}, {Bernstein},
  {Fabricant}, {McCarthy}, \& {Phillips}}]{2004SPIE.5489..441J}
{Johns}, M., {Angel}, J.~R.~P., {Shectman}, S., {et~al.} 2004, in Presented at
  the Society of Photo-Optical Instrumentation Engineers (SPIE) Conference,
  Vol. 5489, Ground-based Telescopes. Edited by Oschmann, Jacobus M., Jr.
  Proceedings of the SPIE, Volume 5489, pp. 441-453 (2004)., ed. J.~M.
  {Oschmann}, Jr., 441--453

\bibitem[{{Kuchner} \& {Traub}(2002)}]{2002ApJ...570..900K}
{Kuchner}, M.~J. \& {Traub}, W.~A. 2002, \apj, 570, 900

\bibitem[{{Macintosh} {et~al.}(2006){Macintosh}, {Graham}, {Palmer}, {Doyon},
  {Gavel}, {Larkin}, {Oppenheimer}, {Saddlemyer}, {Wallace}, {Bauman}, {Evans},
  {Erikson}, {Morzinski}, {Phillion}, {Poyneer}, {Sivaramakrishnan}, {Soummer},
  {Thibault}, \& {Veran}}]{2006SPIE.6272E..18M}
{Macintosh}, B., {Graham}, J., {Palmer}, D., {et~al.} 2006, in Presented at the
  Society of Photo-Optical Instrumentation Engineers (SPIE) Conference, Vol.
  6272, Advances in Adaptive Optics II. Edited by Ellerbroek, Brent L.;
  Bonaccini Calia, Domenico. Proceedings of the SPIE, Volume 6272, pp. 62720L
  (2006).

\bibitem[{{Marois} {et~al.}(2000){Marois}, {Doyon}, {Racine}, \&
  {Nadeau}}]{2000PASP..112...91M}
{Marois}, C., {Doyon}, R., {Racine}, R., \& {Nadeau}, D. 2000, \pasp, 112, 91

\bibitem[{{Mawet} {et~al.}(2005){Mawet}, {Riaud}, {Absil}, \&
  {Surdej}}]{2005ApJ...633.1191M}
{Mawet}, D., {Riaud}, P., {Absil}, O., \& {Surdej}, J. 2005, \apj, 633, 1191

\bibitem[{{Nelson} \& {Sanders}(2006)}]{2006SPIE.6267E..71N}
{Nelson}, J. \& {Sanders}, G.~H. 2006, in Presented at the Society of
  Photo-Optical Instrumentation Engineers (SPIE) Conference, Vol. 6267,
  Ground-based and Airborne Telescopes. Edited by Stepp, Larry M.. Proceedings
  of the SPIE, Volume 6267, pp. 626728 (2006).

\bibitem[{{Racine} {et~al.}(1999){Racine}, {Walker}, {Nadeau}, {Doyon}, \&
  {Marois}}]{1999PASP..111..587R}
{Racine}, R., {Walker}, G.~A.~H., {Nadeau}, D., {Doyon}, R., \& {Marois}, C.
  1999, \pasp, 111, 587

\bibitem[{{Roddier} \& {Roddier}(1997)}]{1997PASP..109..815R}
{Roddier}, F. \& {Roddier}, C. 1997, \pasp, 109, 815

\bibitem[{{Rouan} {et~al.}(2000){Rouan}, {Riaud}, {Boccaletti}, {Cl{\'e}net},
  \& {Labeyrie}}]{2000PASP..112.1479R}
{Rouan}, D., {Riaud}, P., {Boccaletti}, A., {Cl{\'e}net}, Y., \& {Labeyrie}, A.
  2000, \pasp, 112, 1479

\bibitem[{{Sivaramakrishnan} \& {Lloyd}(2005)}]{2005ApJ...633..528S}
{Sivaramakrishnan}, A. \& {Lloyd}, J.~P. 2005, \apj, 633, 528

\bibitem[{{Soummer}(2005)}]{2005ApJ...618L.161S}
{Soummer}, R. 2005, \apjl, 618, L161

\bibitem[{{Soummer} {et~al.}(2003){Soummer}, {Aime}, \&
  {Falloon}}]{2003A&A...397.1161S}
{Soummer}, R., {Aime}, C., \& {Falloon}, P.~E. 2003, \aap, 397, 1161

\bibitem[{{Vernet} {et~al.}(2006){Vernet}, {Kasper}, {V{\'e}rinaud}, {Fedrigo},
  {Tordo}, {Hubin}, {Esposito}, {Pinna}, {Puglisi}, {Tozzi}, {Basden},
  {Goodsell}, {Love}, \& {Myers}}]{2006SPIE.6272E..81V}
{Vernet}, E., {Kasper}, M., {V{\'e}rinaud}, C., {et~al.} 2006, in Advances in
  Adaptive Optics II. Edited by Ellerbroek, Brent L.; Bonaccini Calia,
  Domenico. Proceedings of the SPIE, Volume 6272, pp. 62722K (2006).

\end{thebibliography}
\end{document}